\documentclass{ws-ijmpa}

\begin{document}

\markboth{E. Gourgoulhon, P. Grandcl\'ement \& S. Bonazzola}
{Last orbits of binary black holes}

\catchline{}{}{}

\title{LAST ORBITS OF BINARY BLACK HOLES\footnote{This talk is
dedicated to the memory of our friend and colleague Jean-Alain Marck.}}

\author{\footnotesize \'ERIC GOURGOULHON}

\address{Laboratoire de l'Univers
et de ses Th\'eories, FRE 2462 du C.N.R.S., Observatoire de Paris \\
F-92195 Meudon Cedex, France}

\author{\footnotesize PHILIPPE GRANDCL\'EMENT}

\address{Department of Physics and Astronomy, Northwestern University,
Evanston, IL 60208, USA}

\author{\footnotesize SILVANO BONAZZOLA}

\address{Laboratoire de l'Univers
et de ses Th\'eories, FRE 2462 du C.N.R.S., Observatoire de Paris \\
F-92195 Meudon Cedex, France}

\maketitle

\pub{Received (Day Month Year)}{}

\begin{abstract}
Binary black hole systems in the pre-coalescence stage
are numerically constructed by demanding that the associated
spacetime admits a helical Killing vector.
Comparison with third order post-Newtonian calculations indicates
a rather good agreement until the innermost stable circular orbit.

\end{abstract}

\section{Quasi-equilibrium configurations of binary black holes}	

\subsection{Helical Killing vector}

Binary black holes are considered as one of the
most promising sources
for the gravitational wave interferometric detectors LIGO, VIRGO,
GEO600 and TAMA currently under construction, as well as for the
space interferometer LISA in project.
From the theoretical point of view, computing the final inspiral
and merger of binary black holes has proved to be a very difficult
task. However, when the black
holes are far enough so that the radiation reaction timescale is
large as compared to the dynamical timescale, the orbits can
be approximated as closed ones. Moreover, due to the accumulated effects
of the reaction to gravitational radiation, these orbits can be
considered as circular.

In such a case (closed circular orbits), the
spacetime possesses a one-parameter symmetry group, whose integral curves
are helices. The associated Killing vector is called the {\em helical
Killing vector}. In full general relativity,
requiring the helical symmetry for a binary system imply equal
amounts of outgoing and incoming gravitational radiation.\cite{Detwe89}
The spacetime cannot therefore not be asymptotically flat.\cite{GibboS83}
However, there are two interesting cases where helical symmetry and asymptotic
flatness are compatible: (i) in the post-Newtonian (PN) approximation,
up to the second order and (ii) the Isenberg-Wilson-Mathews (IWM)
approximation of general relativity
(see e.g. Sec. IV.C of Ref.~\cite{FriedUS02}), which assumes that
the spacetime can be foliated by a family of
conformally flat spacelike hypersurfaces and reduces the Einstein
equations to five elliptic equations. Both approximations (i) and (ii)
do not allow for gravitational radiation, which explains why they permit
asymptotically flat spacetimes. They are valid approximations to
general relativity when the gravitational radiation content of
spacetime does not play any important dynamical role. Note that
the IWM formulation contains the first order PN one.

In our first study of the binary black hole problem,\cite{GourgGB02,GrandGB02}
we have used the helical symmetry along with the IWM approximation

\subsection{Rigid rotation}

Regarding the rotation state of the black holes, we have chosen
each black hole to be corotating (synchronized binary), so that the
whole system is in rigid rotation. There are several motivations for
this: (i) It has been recently argued\cite{PriceW01} that when the black
holes are close together, tidal forces may
lead to an efficient transfer of spin angular momentum to orbital
angular, which might lead to synchronization. However this not
clear yet. (ii) As discussed by
Friedman et al.\cite{FriedUS02}, corotating black holes are the only
state compatible with the helical symmetry in full general relativity
(basically, any other rotation state would result in shear and increase
of the horizon area, which is not allowed by the helical symmetry).
Detweiler\cite{Detwe89} has already pointed this by noticing
that gravitational radiation going down in the hole from the
distant past piles up near the distant horizon, resulting in
a divergence of the metric, except for the corotating state.
(iii) For the IWM spacetimes we are considering, other rotation
states are permitted,\cite{Cook02} but the corotating one has
a very simple geometrical interpretation: the black holes
horizons are in this case Killing horizons, i.e. the
helical Killing vector is tangent to the black hole
horizons.\cite{GourgGB02}

\subsection{Determination of the orbital angular velocity}

In the IWM framework, the field equations reduce to five
elliptic equations and the orbital angular velocity $\Omega$
appears only in the boundary condition at infinity
for the corotating shift vector. This contrasts with
the fluid case, where $\Omega$ shows up in the
equation governing the equilibrium of the fluid.
We have determined $\Omega$ by demanding that the ADM
mass be equal to the Komar-like one\cite{GourgGB02},
a requirement which reduces to the classical virial theorem at
the Newtonian limit. We have verified that when the holes
are far apart, this procedure leads to Kepler's third law.

The minimum of ADM mass along a sequence of decreasing
circular orbits with constant horizon area marks
the limit of orbital stability.\cite{FriedUS02}
We call this configuration the ISCO (for {\em
Innermost Stable Circular Orbit}).

\section{Numerical solution}

The five non-linear elliptic equations describing
corotating binary black holes in the IWM approximation
have been solved by means of a multi-domain spectral
method.\cite{BonazGM98,BonazGM99,GrandBGM01}
We use spherical coordinate systems centered on
each black hole, which allow for an easy implementation
of the boundary conditions on the horizons.
Moreover the computational domain extends to spacelike
infinity, thanks to some suitable compactification
of the external domain. This permits a rigorous
implementation of the boundary condition at
infinity (flat spacetime).
The numerical implementation has been performed
by means of an object-oriented code based
on the {\sc Lorene} library.\cite{Lorene}
Numerous tests of the code have been presented
in Ref.\cite{GrandGB02}.

\begin{figure}
\centerline{\psfig{file=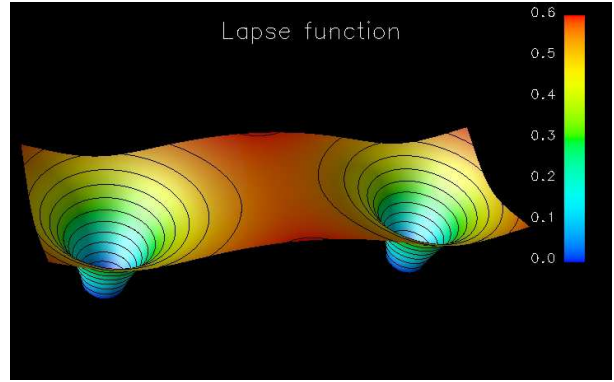,width=8cm}}
\vspace*{8pt}
\caption{\label{f:lapse}
Lapse function in the equatorial plane for the
ISCO configuration.}
\end{figure}

\begin{figure}
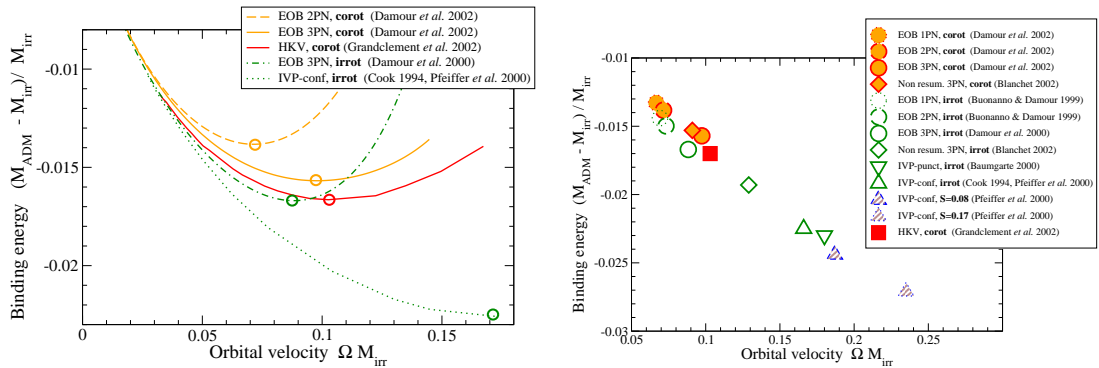

\centerline{\psfig{file=seq_e_omeg.eps,width=7cm} \quad
	    \psfig{file=comp_e_isco.eps,width=7cm}}
\vspace*{8pt}
\caption{\label{f:comp}
Comparison between numerical results and post-Newtonian ones:
{\em left:} binding energy along a constant area sequence of binary black holes.
The numerical results are our helical killing vector approach (HKV)
\protect\cite{GrandGB02}, and
IVP, the initial value approach with effective
potential\protect\cite{Cook94,PfeifTC00};
{\em right:} Values of the binding energy and orbital frequency at the ISCO for
various numerical\protect\cite{Cook94,PfeifTC00,Baumg00,GrandGB02} and
post-Newtonian\protect\cite{DamouGG02,Blanc02,BuonaD99,DamouJS00} methods.}
\end{figure}

\section{Comparison with post-Newtonian results}

Prior to our study\cite{GourgGB02,GrandGB02}, the agreement between numerical
computations of orbiting black holes and analytical (post-Newtonian) ones
was very bad (compare the dotted line with the dot-dashed one in
Fig.~\ref{f:comp}a).
Notably the orbital frequency at the ISCO was differing by a factor of two
(compare the triangles with the circles and diamonds in Fig.~\ref{f:comp}b).
Our results, based on the helical killing vector approach, turned out to
be much closer to the post-Newtonian ones (compare the dark solid line
with the light solid one in Fig.~\ref{f:comp}a, as well as the
square with the circles and diamonds in Fig.~\ref{f:comp}b).
We present a detailed comparison with the Effective One Body (EOB)
analytical approach in Ref.~\cite{DamouGG02}.

\section*{Acknowledgements}

This work has benefited from numerous discussions with
Luc Blanchet, Brandon Carter, Thibault Damour, David Hobill,
J\'er\^ome Novak and Keisuke Taniguchi. We warmly thank all of them.


\begin{thebibliography}{0}

\bibitem{Detwe89}
S.~Detweiler,
in {\em Frontiers in Numerical Relativity},
ed. C.R.~Evans, L.S.~Finn, and D.W.~Hobill
(Cambridge University Press, Cambridge, 1989), p.~43.

\bibitem{GibboS83}
G.W.~Gibbons and J.M.~Stewart,
in {\em Classical General Relativity},
ed. W.B.~Bonnor, J.N.~Islam and M.A.H.~MacCallum
(Cambridge University Press, Cambridge, England, 1983), p.~77.

\bibitem{FriedUS02}
J.L.~Friedman, K.Uryu, and M.~Shibata,
{\it Phys. Rev. D} {\bf 65}, 064035 (2002).

\bibitem{GourgGB02}
E.~Gourgoulhon, P.~Grandcl\'ement, and S.~Bonazzola,
{\it Phys. Rev. D} {\bf 65}, 044020 (2002).

\bibitem{GrandGB02}
P.~Grandcl\'ement, E.~Gourgoulhon, and S.~Bonazzola,
{\it Phys. Rev. D} {\bf 65}, 044021 (2002).

\bibitem{PriceW01}
R.H.~Price and J.T.~Whelan,
{\it Phys. Rev. Lett.} {\bf 87}, 231101 (2001).

\bibitem{Cook02}
G.B.~Cook,
{\it Phys. Rev. D} {\bf 65}, 084003 (2002).

\bibitem{BonazGM98}
S.~Bonazzola, E.~Gourgoulhon, and J.-A.~Marck,
{\it Phys. Rev. D} {\bf 58}, 104020 (1998).

\bibitem{BonazGM99}
S.~Bonazzola, E.~Gourgoulhon, and J.-A.~Marck,
{\it J. Comput. Appl. Math.} {\bf 109}, 433 (1999).

\bibitem{GrandBGM01}
P.~Grandcl\'ement, S.~Bonazzola, E.~Gourgoulhon, and J.-A.~Marck,
{\it J. Comp. Phys.} {\bf 170}, 231 (2001).

\bibitem{Lorene}
{\tt http://www.lorene.obspm.fr}

\bibitem{Cook94}
G.B.~Cook,
{\it Phys. Rev. D} {\bf 50}, 5025 (1994).

\bibitem{PfeifTC00}
H.P.~Pfeiffer, S.A.~Teukolsky, and G.B.~Cook,
{\it Phys. Rev. D} {\bf 62}, 104018 (2000).

\bibitem{Baumg00}
T.W.~Baumgarte,
{\it Phys. Rev. D} {\bf 62}, 024018 (2000).

\bibitem{DamouGG02}
T.~Damour, E.~Gourgoulhon, and P.~Grandcl\'ement,
{\it Phys. Rev. D}, submitted (preprint: gr-qc/0204011).

\bibitem{Blanc02}
L.~Blanchet,
{\it Phys. Rev. D}, in press (gr-qc/0112056).

\bibitem{BuonaD99}
A.~Buonanno and T.~Damour,
{\it Phys. Rev. D} {\bf 59}, 084006 (1999).

\bibitem{DamouJS00}
T.~Damour, P.~Jaranowski, and G.~Sch\"afer
{\it Phys. Rev. D} {\bf 62}, 084011 (2000).

\end{thebibliography}
\end{document}